# High Performance Inverted Organic Photovoltaics Without Hole Selective Contact.

*Achilleas Savva\*, Ignasi Burgués-Ceballos, Giannis Papazoglou and Stelios A. Choulis\*.*

Molecular Electronics and Photonics Research Unit, Department of Mechanical Engineering and Materials Science and Engineering, Cyprus University of Technology, Limassol, 3603 (Cyprus).

\*email: achilleas.savva@cut.ac.cy, \*email: stelios.choulis@cut.ac.cy

**ABSTRACT**

A detailed investigation of the functionality of inverted organic photovoltaics (OPVs) using bare Ag contacts as top electrode is presented. The inverted OPVs without hole transporting layer (HTL) exhibit a significant gain in hole carrier selectivity and power conversion efficiency (PCE) after exposure in ambient conditions. Inverted OPVs comprised of ITO/ZnO/poly(3-hexylthiophene-2,5-diyl):phenyl-C61-butyric acid methyl ester (P3HT:PCBM)/Ag demonstrate over 3.5% power conversion efficiency only if the devices are exposed in air for over 4 days. As concluded through a series of measurements, the oxygen presence is essential to obtain fully operational solar cell devices without HTL. Moreover, accelerated stability tests under damp heat conditions (RH=85% and T=65$^o$C) performed to non-encapsulated OPVs demonstrate that HTL-free inverted OPVs exhibit comparable stability to the reference inverted OPVs. Importantly, it is shown that bare Ag top electrodes can be efficiently used in inverted OPVs using various high performance polymer:fullerene bulk heterojunction material systems demonstrating 6.5% power conversion efficiencies.

# 1. INTRODUCTION

Solution-based thin film photovoltaics based on organic semiconducting materials have attracted remarkable interest as a possible alternative to conventional, inorganic photovoltaic technologies. Newly synthesized organic semiconductors, as well as novel interfacial engineering and electrode configurations continuously push power conversion efficiency (PCE) of these photovoltaic devices over 10%.[1-2]

Typically, in all aforementioned device structures, the absorber is sandwiched between two electrodes, each one selectively extracting one type of charge carrier. The selectivity of the electrodes is a crucial factor for high device performance and is typically provided by the implementation of sophisticated electrodes comprised by charge selective contacts between the absorber and the metallic terminals of the device. The normal structure is usually based on: ITO/poly(3,4-ethylene dioxythiophene):poly(styrenesulfonate) (PEDOT:PSS)/photoactive layer/low work function (LWF) metal (i.e. Al).[3] In inverted structured OPVs the current flow is reversed by changing the polarity of the electrodes and is normally based on: ITO/ n-type metal oxides (i.e. $TiO_x$[4] or ZnO[5] or n-doped metal oxides[6])/photoactive layer/high work function (HWF) metal (i.e. Ag)

Lifetime is an equally important factor relevant to product development targets of OPVs. It has been proven that electrodes are one of the main origins of failure of OPVs under harsh environmental conditions.[7] The use of LWF metal based cathodes is a main degradation factor related to electrodes stability, since LWF metals (Li, Ca, Al) are oxidized extremely fast. Several strategies to improve the stability of normal structured OPVs have been proposed such as the addition of metal nanoparticles within the active layer[8-9] as well as the implementation of more stable interfacial layers.[10-11]

On the other hand, inverted solution based PVs provide a facile and reliable strategy to improve OPV stability, due to the implementation of HWF, more stable metals (i.e. Ag) as top electrode.[12] Despite the enhanced lifetime of the inverted structured compared with normal structured OPVs, it has been proven that a major origin of failure of inverted OPVs is due to the most commonly used hole selective contact PEDOT:PSS.[13] Its hygroscopic and acidic nature results in insufficient hole selectivity of the top electrode over time of exposure under harsh environmental conditions.[14-15] In addition, we have recently proved that not only the hygroscopic nature of PEDOT:PSS, but also the poor adhesion between the PEDOT:PSS and the polymeric active layer materials of inverted OPVs is another mechanism of degradation of inverted OPVs under intense humidity conditions.[16] In addition, a number of studies prove that major degradation mechanisms of inverted OPVs arise from the interfaces formed between the top electrode components and the active layer when PEDOT:PSS is used.[17-18]

Based on the latter, promising replacements and currently investigated substitutes for PEDOT:PSS come from the class of metal oxides, due to their excellent optoelectronic properties and chemical/moisture resistance. Metal oxides such as $WO_3$,[19] $MoO_3$,[20] $V_2O_5$,[21] have been used as efficient hole selective contacts in inverted OPVs. This novel buffer layer engineering, results in optimum inverted OPVs electrodes selectivity, leading in high Fill Factors (FF) over 65%,[22] and in some cases in enhanced lifetime performance.[23]

Interestingly, recent studies report that the electrodes of inverted OPVs could provide the necessary charge selectivity without the use of charge selective contacts. J.-C.Wang *et al.* reported an efficient inverted OPV device in which an electron selective layer was not used.[24] The inverted OPVs with bare ITO bottom electrodes demonstrated high electron selectivity after a UV light treatment. In addition, M.S. white *et al.* demonstrated that inverted OPVs using bare Ag top contacts could efficiently serve as hole selective electrode for inverted OPVs.[25] Further studies analyzed the phenomenon and observed that inverted OPVs showed an increase in PCE after exposure of the devices in air.[26] The increase in PCE was attributed to the increased work function of Ag layers after the exposure in air. TOF-SIMs studies showed the presence of significantly increased silver oxide percentage at the interface of Ag with the P3HT:PCBM layer.[27]

In this report a detailed investigation of the functionality of inverted OPVs using bare Ag contacts as the hole selective top electrode is provided. Initially, inverted OPVs comprised of ITO/ZnO/P3HT:PCBM/PEDOT:PSS/Ag (reference inverted OPVs) are compared with ITO/ZnO/P3HT:PCBM/Ag (HTL-free inverted OPVs). These devices are measured after exposure in air for several days after fabrication. It is demonstrated that HTL-free devices are continuously gaining in PCE and finally reaching the PCE levels of the reference inverted OPVs. To investigate the impact of atmospheric conditions in the hole selectivity process encapsulated and non-encapsulated HTL-free devices are compared during several days after exposure in air. The oxygen presence is essential to obtain fully operational solar cell devices. This effect is verified through a series of measurements and calculations such as current vs voltage characteristics, Vbi calculations, statistical analysis and photocurrent mapping measurements. Accelerated stability tests under damp heat conditions (RH=85% and T=65ºC) performed in non-encapsulated devices demonstrated that HTL-free inverted OPVs exhibit comparable stability compared with reference inverted OPVs (using PEDOT: PSS) at least for the first 200 hours of testing. Finally, bare Ag top electrodes under the presence of suitable oxygen treatment are proved to be functional using different high performance polymer-fullerene active layer material systems such as PTB7:PC[70BM], and PTB7-TH:PC[70]BM demonstrating HTL-free inverted OPVs with 6.5% PCE .

## 2. MATERIALS AND METHODS

*Materials:* Pre-patterned glass-ITO substrates (sheet resistance 4Ω/sq) were purchased from Psiotec Ltd. Zinc acetate dehydrate, 2-methoxyethanol and ethanolamine have been purchased from Sigma Aldrich, P3HT from Rieke metals, PTB7 from 1-material, PTB7-Th from Solarmer, PC[60]BM and PC[70]BM from Solenne BV and PEDOT:PSS PH from H.C. Stark. *Device Fabrication:* For inverted solar cells, ITO substrates were sonicated in acetone and subsequently in isopropanol for 10 minutes. ZnO electron transporting layer was prepared using a sol-gel process as described in detail in our previous study.[6] The photo-active layer, deposited on top of ZnO, consisted of a) a blend of P3HT:$PC_{60}$BM (1:0.8 wt%) 36 mg/ml in chlorobenzene, doctor bladed in air, with a resulting thickness of ~180 nm, b) a blend of PTB7:$PC_{70}$BM (1:1.5) 25 mg/mL in chlorobenzene with 3% of 1,8-diiodooctane (DIO) additive, doctor bladed in air with a resulting thickness of ~90 nm without further annealing, or c) a blend of PTB7-Th:$PC_{70}$BM (1:2) 36 mg/mL in o-dichlorobenzene with 2.5% of DIO additive, spin coated in glove box, ~90 nm thick, slow dried in a petri dish for 1h. For inverted OPVs containing PEDOT:PSS a treatment with two wetting agents was applied as described in detail previously.[28] All the inverted OPVs based on P3HT:PCBM were annealed inside a glove box at 140 °C for 20 minutes. The devices were completed by thermal evaporating a silver layer with a thickness of 100 nm. Encapsulation was applied directly after evaporation in the glove box using an Ossila E131 encapsulation epoxy resin activated by 365nm UV-irradiation and a glass coverslip. The active area of the devices was 9 $mm^2$. *Storage*: During the study the samples were stored under two different conditions, namely 1) exposure to air and 2) exposure to pure oxygen. For the latter the samples were stored in a desiccator, to which subsequential vacuum and refilling with pure oxygen (99.5%) were performed. *Accelerated degradation*: The un-encapsulated inverted OPVs were subjected to degradation under the ISOS D-3 protocol (Damp Heat test, RH = 85%, T = 65 °C, Dark conditions) using a climate chamber. *Characterization:* The thicknesses of the active layers were measured with a Veeco Dektak 150 profilometer. The current density-voltage (J/V) characteristics were measured with a Keithley source measurement unit (SMU 2420). For illumination, a calibrated Newport Solar simulator equipped with a Xe lamp was used, providing an AM1.5G spectrum at 100mW/cm$^2$ as measured by an oriel 91150V calibration cell equipped with a KG5 filter. Net photocurrent vs voltage characteristics were obtained by extracting the dark from the illuminated J/V characteristics. Photocurrent and open circuit voltage (Voc) mapping measurements were performed under 405 nm laser excitation using a Botest PCT photocurrent system.

## 3. RESULTS AND DISCUSSION

Inverted OPVs comprised of ITO/ZnO/P3HT:PCBM/ PEDOT:PSS/Ag (reference inverted OPVs) are compared with ITO/ZnO/ P3HT:PCBM/Ag (HTL-free inverted OPVs) and shown in figure 1a. The inverted OPVs under study were tested directly after fabrication with no encapsulation barrier, stored in ambient conditions and tested periodically up to 7 days after fabrication. To avoid repetitions within the rest of the manuscript the process of exposing in ambient conditions of all the inverted OPVs under study 2, 4 and 7 days after fabrication will be referred as day 2, 4 and 7, respectively. Figure 1 b) c) and d) show representative J/V characteristics for the inverted OPVs under study out of total 8 inverted OPV devices in each case. Similar results were observed in more than 5 identically executed experimental runs (over 40 devices for each case).

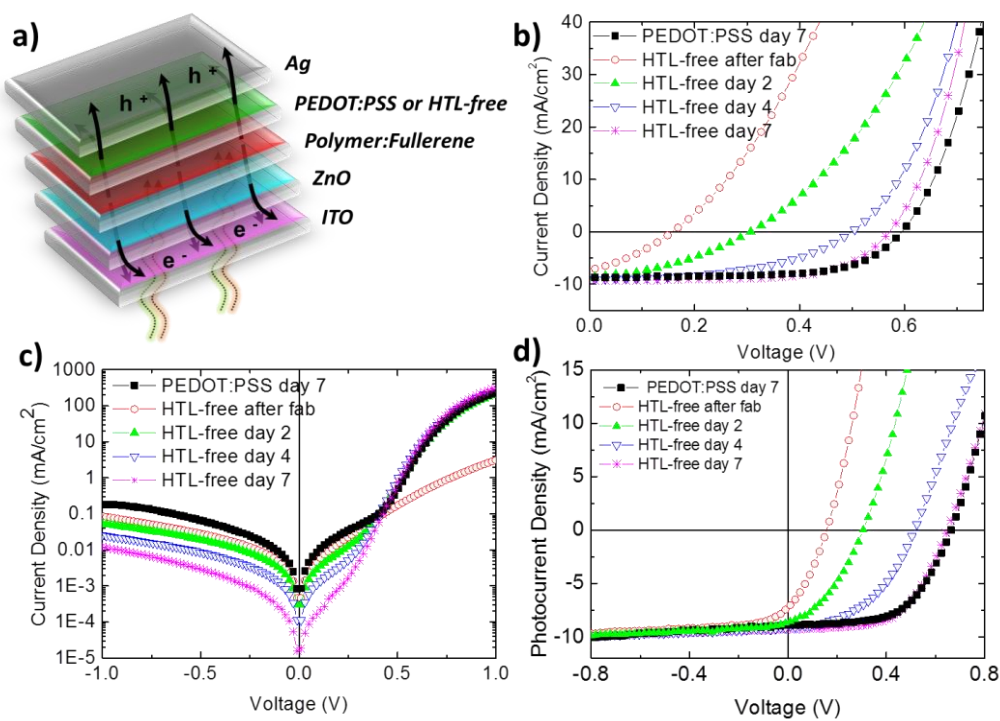

**Figure 1. a)** the inverted OPVs structure and materials used in this study, **b)** current density versus voltage characteristics under illumination and **c)** under dark conditions, **d)** net photocurrent density versus voltage measurements.

Figure 1b shows the current density vs voltage characteristics of all of the non-encapsulated inverted OPVs under study obtained from day 0 to day 7. Reference inverted OPVs exhibited good cell operation directly after fabrication. This good performance was also maintained upon exposure of reference devices in air after 2, 4 and 7 days, respectively. On the other hand, non-encapsulated HTL-free inverted OPVs exhibited poor device performance after fabrication with low Voc and FF factor values and thus limited initial PCE values. Interestingly, these HTL-free devices were continuously gaining in Voc, FF and PCE after the exposure of the devices in air and finally reaching the PCE levels of the reference inverted OPVs.

These observations are also in agreement with figure 1c, in which the J/V characteristics under dark conditions for all the inverted OPVs under study are shown. Functionality of non-encapsulated HTL-free inverted OPVs improved gradually from day 0 to day 7. This enhancement could be mainly attributed to a drastic decrease of the series resistance (Rs) and secondary to parallel resistance (Rp) increase from day 0 to day 7 respectively. At day 7 non-encapsulated HTL-free inverted OPVs show good hole selectivity since the device internal resistances, FF and PCE exhibit good values, very similar to the reference inverted OPVs. The latter, clearly demonstrates a functional hole selective top electrode for inverted OPVs, using only Ag without HTL. The values of Rs and Rp for all the representative diodes under study were calculated using a simulation model described previously by Waldauf et al[29] and shown in table 1.

In an attempt to better understand the origin of the functionality of the bare Ag electrodes of inverted OPVs, the Vbi of all the inverted OPVs under study were calculated. Figure 1d shows the net photocurrent density as a function of diode bias for the three inverted OPVs under study. These measurements can be used to determine the built-in potential (Vbi) and thus the changes in the energy barriers at the interfaces between the active layer and the electrodes.[30] As table 1 shows, the Vbi for HTL-free inverted OPVs increased gradually over exposure in air from 0.16 V after fabrication to 0.52 V in day 4 and 0.65 V in day 7. The Vbi of HTL-free inverted OPVs at day 7 is very similar with the corresponding 0.67 V of the reference inverted OPVs using PEDOT:PSS hole selective contact directly after fabrication. All the critical device parameters of the representative inverted OPVs under study are shown in table 1.

Table 1. Summary of the photovoltaic parameters of all the inverted OPVs under study calculated from figure 1.

| Inverted OPVs | $V_{oc}$ [V] | $J_{sc}$ [mA/cm$^2$] | FF [%] | PCE [%] | $R_p$ [Ohm] | $R_s$ [Ohm] | $V_{bi}$ [V] |
|---|---|---|---|---|---|---|---|
| Reference after fab. | 0.59 | 8.71 | 64.5 | 3.34 | 2795 | 0.62 | 0.67 |
| HTL-free after fab. | 0.16 | 7.25 | 32.7 | 0.37 | 470 | 150 | 0.16 |
| HTL-free at Day 2 | 0.31 | 8.64 | 37.5 | 0.99 | 1271 | 1.16 | 0.31 |
| HTL-free at Day 4 | 0.50 | 9.10 | 48.0 | 2.18 | 1558 | 1.0 | 0.52 |
| HTL-free at Day7 | 0.57 | 9.33 | 64.8 | 3.45 | 1570 | 1.1 | 0.65 |

From day 1 to 7 the most important changes in the device are the Voc, Vbi and the Rs. It could be deduced that at day 0 an energy barrier at the P3HT:PCBM/Ag interface is present, according to the high Rs and low Vbi. Over days of exposure this barrier is reduced, leading to continuously lower Rs, higher Vbi and Voc and thus increased hole selectivity, FF and PCE values. The origin of this barrier is attributed to a reduction of Ag work function. The work function of Ag has been documented to be located at −4.3 eV.[31] However, it has been shown that exposure to oxygen can induce a shift in band alignment at metal/organic interfaces.[31] Based on the efficient hole-collecting nature of the P3HT/Ag interface seen here, it is evident that a similar shift further from vacuum is occurring upon contact of the two materials with the presence of oxygen.

To proof that oxygen is the only component causing the modification of the silver electrode we compared HTL-free inverted OPVs exposed to ambient air with HTL-free inverted OPVs exposed to >99% oxygen atmosphere (supplementary figure S1). We observed that HTL-free inverted OPVs exposed to oxygen only environment became functional at day 7, following the same trend as those exposed to ambient air.

This suggests that the presence of oxygen is crucial in this process. According to our results, water and other gases present might play a minor role.

Although Ag band alignment with the P3HT interface could serve as a hole transporting material, it would be unlikely to also provide electron blocking capabilities and thus the high FF values observed in HTL-free devices is most likely related to changes in interfacial properties. It has been previously reported that upon air exposure, there is a pronounced increase of AgO and $Ag_2O$ signals at the Ag–organic interface.[27] The presence of an oxide layer at the Ag–organic interface in the samples exposed to oxygen can be used to justify the outstanding hole selectivity of the inverted OPVs with no hole selective layer.

To further investigate the impact of oxygen in this process, inverted OPVs with and without encapsulation were fabricated. Four series of inverted OPVs were tested in this experimental run. Reference and HTL-free inverted OPVs were fabricated and tested up to 7 days after fabrication, similar to what is described previously in this study. Some of the HTL-free devices were encapsulated directly after fabrication in nitrogen filled glovebox before the devices were exposed to oxygen (named as HTL-free encapsulated after fab). Another set of HTL-free inverted OPVs were encapsulated after being exposed 2 days in ambient conditions (HTL-free encapsulated after 2 days). Figure 2 shows the box plots constructed out of 8 devices for each inverted OPV device structure under study.

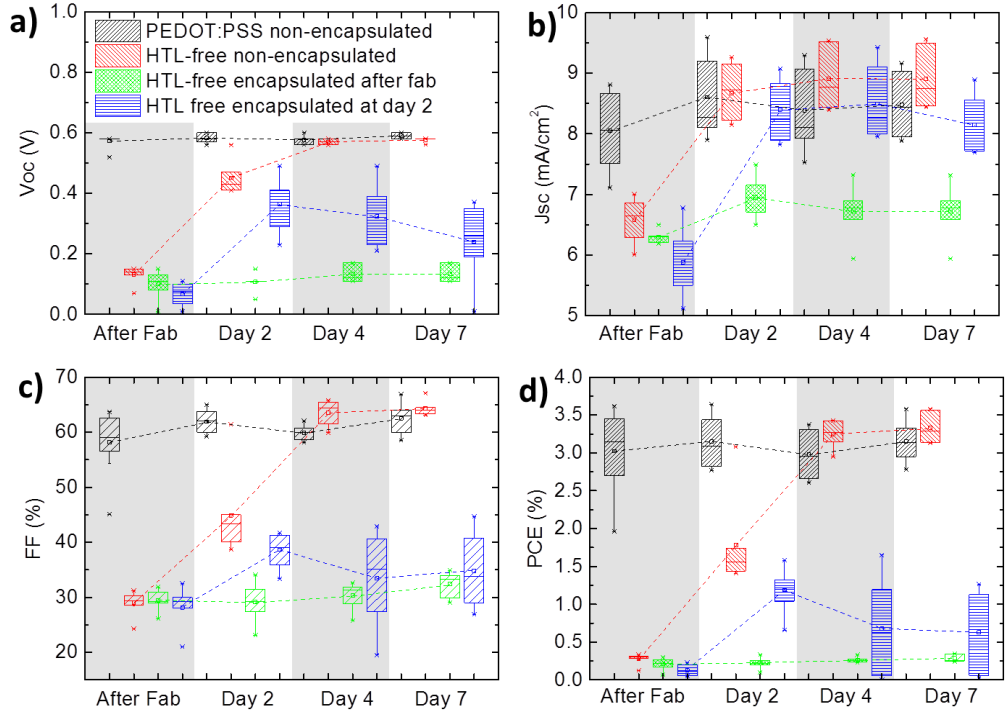

**Figure 2.** Average results represented in box plots out of 8 devices of each of the 4 series of inverted OPVs under study, reference inverted OPVs (black box plots), non-encapsulated HTL-free inverted OPVs (red box plots), encapsulated after fabrication HTL-free inverted OPVs (green box plots) and encapsulated 2 days after fabrication and exposure in air HTL-free inverted OPVs (blue box plots). **a)** Open circuit voltage (Voc) **b)** current density (Jsc) **c)** fill factor (FF) and **d)** power conversion efficiency (PCE).

Reference inverted OPVs with PEDOT:PSS hole selective contact demonstrated good functionality directly after fabrication and well maintained up to day 7. On the other hand, all the HTL-free un-encapsulated inverted OPVs demonstrated very low Voc, FF and PCE values after fabrication. Under ambient exposure these devices showed a significant increase mainly in Voc. FF parameter is also improved and thus PCE values are greatly increased. In contrast, HTL-free OPVs encapsulated after fabrication did not gain in PCE from day 1 to day 7, exhibiting very low Voc, FF and PCE values. This is a strong indication that the presence of oxygen is necessary for gaining in hole selectivity and thus PCE. As a further confirmation to the above effects, HTL-free inverted OPVs encapsulated 2 days after fabrication exhibited an increase in the first 2 days (when atmospheric oxygen is present) but after encapsulation this gaining stops and the PCE is "frozen" at the values measured just before the encapsulation process in day 2. This observation proves that the oxygen presence is necessary for over 5 days in order to obtain fully operational inverted OPVs using bare Ag hole selective electrodes. It is worthy to note here that similar results have been observed in several other experimental runs. In addition, we observed that non-encapsulated devices stored in a nitrogen filled glovebox (instead of ambient conditions) did not convert into fully operational devices even after several days (devices were periodically measured but data are not shown within the manuscript). This is another proof that the oxygen presence is a crucial factor in functionalizing the hole selective electrode.

In order to examine whether the observed effect is reversible, 8 HTL-free devices fully functional (after exposure to air) were subjected to a $10^{-3}$ bar vacuum for 3 hours. As demonstrated in supplementary figure S2b the inverted OPVs do not loose in functionality after exposure under vacuum. This indicates that the observed effect is not reversible. In addition, another set of 8 HTL-free inverted OPVs were reversed engineered by removing the silver layer on top after the inverted OPVs became fully functional at day 7. After that, a fresh layer of Ag was evaporated on top of the ITO/ZnO/P3HT:PCBM. The aforementioned HTL-free inverted OPVs exhibited similar diode behavior with that of day 0 (low Voc and PCE). These results (see figure S2b in supplementary information) indicate that the observed effect originates from the Ag layer interaction with oxygen and not from any other interaction between the layers of the device.

To better analyze the phenomenon, spatially resolved Voc and photocurrent measurements were performed over the whole area of all the inverted OPVs under study. Figure 3 shows the Voc maps of the representative reference inverted OPVs directly after fabrication and HTL-free inverted OPVs after fabrication, at day 2 and at day 7.

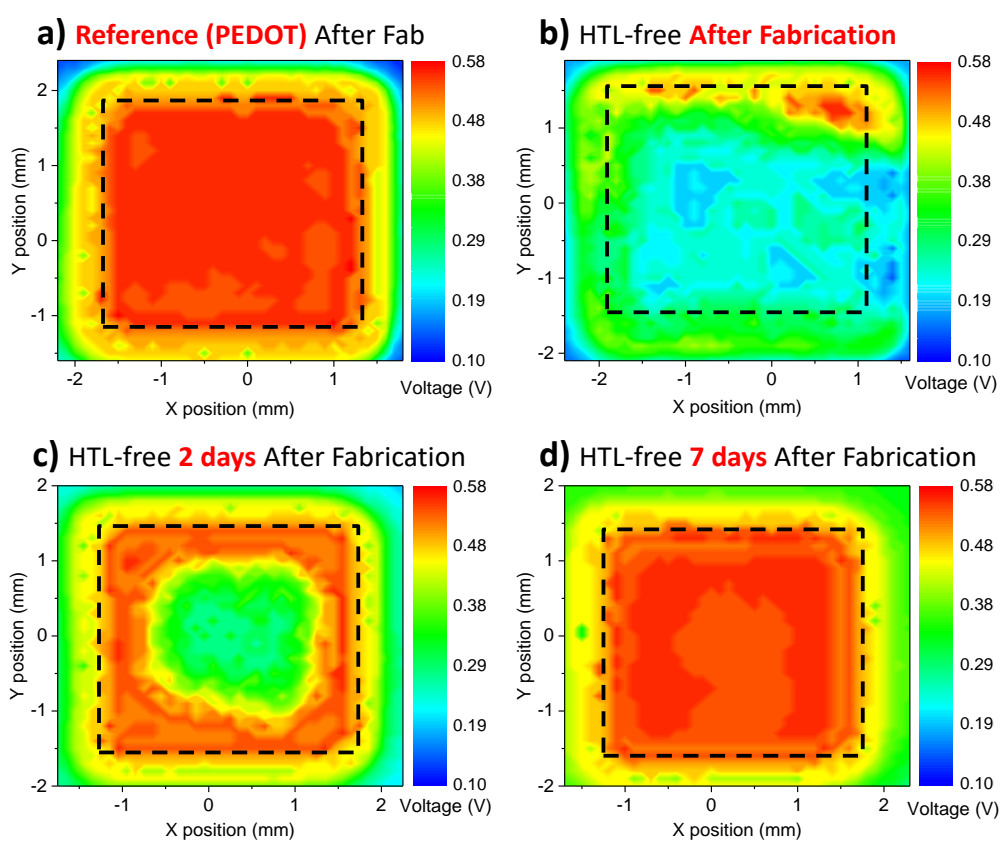

**Figure 3.** Voc maps at 405nm laser excitation of **a)** reference inverted OPVs after fabrication, **b)** non-encapsulated HTL-free inverted OPVs after fabrication, **c)** non-encapsulated HTL-free inverted OPVs 2 days after fabrication and **d)** non-encapsulated HTL-free inverted OPVs 7 days after fabrication. The active area of all the inverted OPVs under study is 9 mm$^2$ and is highlighted with a dashed black square.

Reference inverted OPVs (figure 3a) showed a well distributed Voc within the device directly after fabrication. Interestingly, HTL-free inverted OPVs after fabrication showed a poor Voc only at one edge of the device. Testing the same inverted OPV at day 2 revealed an intense Voc at the edges and moving to the center of the device. Finally, at day 7 the Voc is well distributed all over the 9 mm$^2$ of the device. Consistently, photocurrent maps (see Fig. S3 in supplementary information) follow a similar trend as Voc. These measurements undoubtedly demonstrate that the process takes place from the edges to the center of the HTL-free devices. We believe that this is linked with our previous observations concerning the impact of oxygen in the HTL-free inverted OPVs. The oxygen presumably penetrates the non-encapsulated inverted OPVs from the edges of the device and the Voc and photocurrent are correspondingly more intense at the edges at day 0 and 2. At day 7 the oxygen has diffused all over the active area, causing a reduction in Ag work function and correspondingly a homogeneous Voc and an efficient photocurrent generation due to enhanced top electrode hole selectivity. These observations are in agreement with the assumptions made in previously reported studies that the oxygen might diffuse from the sides of the Ag electrodes rather than through the Ag layer.[26]

Another important factor for cost efficient OPVs is their long-term stability. It is well known that electrodes are one of the major degradation mechanisms of inverted OPVs.[9a] The inverted OPVs without any encapsulation barrier were subjected to stability studies under the ISOS D-3 protocol using a climate chamber. Damp Heat test (RH = 85% - T = 65 °C –Dark) is considered as one of the harshest test for OPVs and it has been found to mainly affect the electrodes of inverted OPVs.[7] Our reference inverted OPVs (ITO/ZnO/P3HT:PCBM/PEDOT:PSS/Ag) were compared with ITO/ZnO/P3HT:PCBM/Ag (HTL-free inverted OPVs). 12 non-encapsulated devices in each case were examined. Figure 4 shows the average results of the normalized Voc, Jsc, FF and PCE over time of damp heat exposure.

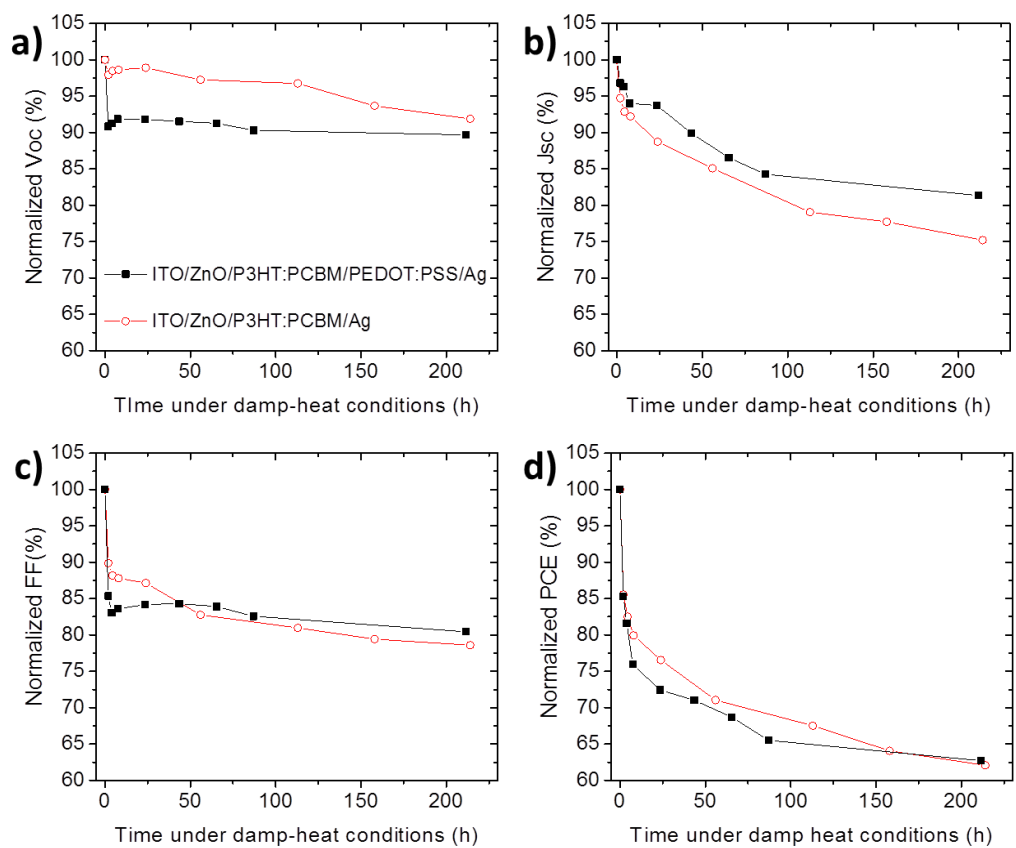

**Figure 4.** Lifetime performance under ISOS D-3 protocol (Damp heat, RH = 85%-Heat 65 °C -Dark) for reference non-encapsulated inverted OPVs ITO/ZnO/P3HT:PCBM/ PEDOT:PSS/Ag (black filled squares) and HTL-free non-encapsulated inverted OPVs ITO/ZnO/P3HT:PCBM/Ag (open red circles). **a)** Normalized Voc, **b)** Normalized Jsc, **c)** Normalized FF and **d)** Normalized PCE over time of exposure under damp-heat conditions.

HTL-free inverted OPVs under damp heat conditions exhibited comparable lifetime performance with inverted OPVs using a PEDOT:PSS. Both of the compared inverted OPVs exhibit a significant drop in Jsc, FF and PCE in the first few hours. Later on, the drop is smoother in both cases until the end of the study. Therefore, initial damp heat tests prove that HTL-free inverted OPVs exhibit lifetime performances comparable with inverted OPVs using a PEDOT:PSS as hole selective layers. However, a more detailed lifetime investigation would be desirable in order to better examine the lifetime behavior of these devices for more than 200 hours. To identify the exact degradation mechanisms in each case deserves further studies and is beyond the scope of the present work.

Finally, in order to examine the universality of this phenomenon, different polymer:fullerene photo-active layer systems were tested. Two high performing donor conjugated polymers were used, namely poly({4,8-bis[(2-ethylhexyl)oxy]benzo[1,2-b:4,5-b′]dithiophene-2,6-diyl}{3-fluoro-2-[(2-ethylhexyl)carbonyl]thieno[3,4-b]thiophenediyl}) (PTB7) and poly[4,8-bis(5-(2-ethylhexyl)thiophen-2-yl)benzo[1,2-b;4,5-b']dithiophene-2,6-diyl-alt-(4-(2-ethylhexyl)-3-fluorothieno[3,4-b]thiophene-)-2-carboxylate-2-6-diyl)] (PBDTT-FTTE, *aka* PTB7-Th). The same device structures were compared (with and without $MoO_3$ as HTL) similarly as describe before, and all the devices were systematically tested over 7 days of exposure in ambient conditions. Representative J/V plots of all the inverted OPVs under study are shown in Figure 5. In total, 12 devices of each type were tested.

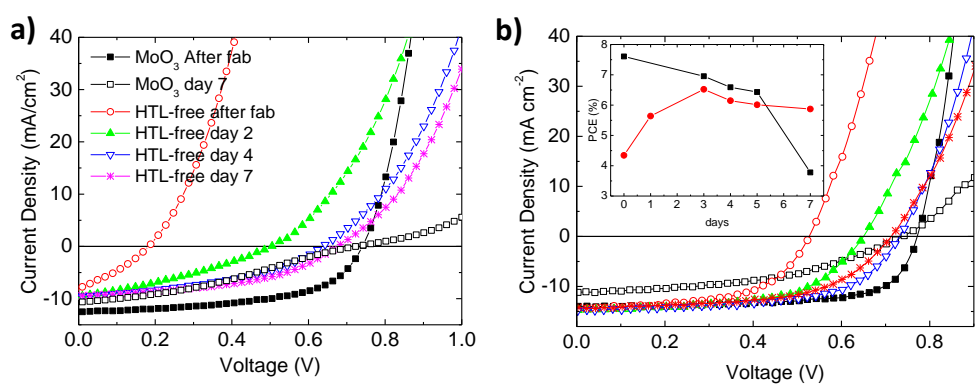

**Figure 5.** Current density vs voltage characteristics under illumination of OPVs based on a) PTB7:PC[70]BM and b) PTB7-Th:PC[70]BM with and without HTL obtained at different days of exposure in air. Inset in b) shows the evolution of PCE in devices with (black squares) and without (red circles) HTL.

In the two material systems, reference inverted OPVs (with $MoO_3$) exhibited good diode behavior after fabrication demonstrating Voc = 0.74 V and 0.77 V, Jsc = 12.6 mA/cm$^2$ and 14.1 mA/cm$^2$, FF = 63% and 70% and PCE = 5.9% and 7.6% for PTB7 and PTB7-Th based OPVs, respectively. On the other hand, HTL-free inverted OPVs exhibited very poor performance after fabrication, similarly to what was observed when using our reference P3HT:PCBM as the photo-active layer material system. Over several days in exposure to air, HTL-free devices showed a gradual increase in all the photovoltaic parameters. After 7 days, PTB7-based HTL-free devices reached Voc = 0.68 V, Jsc = 9.5 mA/cm$^2$, FF = 60% and PCE = 3.9%. In the case of PTB7-Th HTL-free devices, only 3 days of exposure to air were needed to increase efficiency from the initial 4.3% to a respectable PCE of 6.5%, with Voc = 0.73 V, Jsc = 14.9 mA/cm$^2$ and FF = 60%. In both cases, equivalently as in P3HT:PCBM samples, the greatest enhancement corresponds to a significant increase in Voc, as can be clearly seen in Figure 5. This indicates a reduction in Ag work function, as analyzed previously within the text. Therefore, this phenomenon seems to be independent of the photo-active materials. On the contrary, the PCE values of the HTL-free OPVs based on these two high performing conjugated polymer donors did not match those of the reference device. Inset in Figure 5b shows a constant decrease in the performance of reference devices during the first 5 days, followed by a more abrupt decrease from there on. We attribute the former to a gradual degradation of the photoactive layer, as these materials are known to be not particularly air stable.[32] The second step could be related to $MoO_3$ degradation,[33] which we have systematically observed in other material systems containing $MoO_3$ (this will be published separately). As a result, the progressive increase of PCE in HTL-free devices upon exposure to air competes with the gradual degradation of the active layer, as inset in Figure 5b reveals. Based on these observations, we believe that bare Ag could serve as an efficient hole selective electrode for a broad variety of active layer systems, particularly for those that show high air stability.

## 4. CONCLUSIONS

In summary, functional inverted OPVs without hole selective contact are investigated. It is shown that a crucial requirement to achieve sufficient hole selectivity for inverted OPVs using bare Ag top electrodes is the exposure of the devices in ambient conditions for several few days. As proven by detailed J/V analysis, the inverted OPVs without hole selective contact are gaining gradually in Voc, FF and PCE after exposure in ambient conditions. In contrast, HTL-free inverted OPVs which have been encapsulated before any contact with atmospheric conditions do not convert into operational diodes even after 7 days of exposure. Specifically, it is demonstrated that the presence of oxygen is a key factor to achieve the desired hole selectivity and that the process takes place from the edges to the center of the device. Progressive oxygen ingress is assumed to partially oxidize the Ag layer in the interface with the active layer, inducing changes in the work function of the electrode. Furthermore, damp heat test for the air stable P3HT:PCBM material system proved that HTL-free non-encapsulated inverted OPVs exhibit comparable lifetime with inverted non-encapsulated OPVs using PEDOT:PSS/Ag hole selective electrode at least up to T80. Importantly, the universality of this effect is demonstrated: bare Ag could be served as functional hole selective electrodes not only in the P3HT:PCBM case but also in other highly efficient polymer:fullerene systems such as PTB7:PCB[70]BM and PTB7-Th:PC[70]BM leading in HTL-free inverted OPVs with PCE of 6.5%. We believe that the concept and detailed investigation for HTL-free OPVs presented could be used for the evaluation of air-stable novel materials simplifying the processing steps for high performance inverted OPVs.

AKNOWLEDGEMENTS

This work has been funded by the H2020-ERC-2014-GoG project *"Solution Processed Next Generation Photovoltaics (Sol-Pro)"* number 647311.

REFERENCES

1. Guo, X.; Zhou, N.; Lou, S. J.; Smith, J.; Tice, D. B.; Hennek, J. W.; Ortiz, R. P.; Navarrete, J. T. L.; Li, S.; Strzalka, J.; Chen, L. X.; Chang, R. P. H.; Facchetti, A.; Marks, T. J., Polymer Solar Cells With Enhanced Fill Factors. *Nat. Photonics* **2013,** *7,* 825-833.

2. You, J.; Dou, L.; Yoshimura, K.; Kato, T.; Ohya, K.; Moriarty, T.; Emery, K.; Chen, C.-C.; Gao, J.; Li, G.; Yang, Y., A Polymer Tandem Solar Cell With 10.6% Power Conversion Efficiency. *Nat. Commun.* **2013,** *4*, 1446.

3. Bai, S.; Cao, M.; Jin, Y.; Dai, X.; Liang, X.; Ye, Z.; Li, M.; Cheng, J.; Xiao, X.; Wu, Z.; Xia, Z.; Sun, B.; Wang, E.; Mo, Y.; Gao, F.; Zhang, F., Low-Temperature Combustion-Synthesized Nickel Oxide Thin Films as Hole-Transport Interlayers for Solution-Processed Optoelectronic Devices. *Adv. Energy Mater.* **2014,** *4*, 1301460.

4. Waldauf, C.; Morana, M.; Denk, P.; Schilinsky, P.; Coakley, K.; Choulis, S. A.; Brabec, C. J., Highly Efficient Inverted Organic Photovoltaics Using Solution Based Titanium Oxide as Electron Selective Contact. *Appl. Phys. Lett.* **2006,** *89*, 233517.

5. Oh, H.; Krantz, J.; Litzov, I.; Stubhan, T.; Pinna, L.; Brabec, C. J., Comparison of Various sol–gel Derived Metal Oxide Layers for Inverted Organic Solar Cells. *Sol. Energy Mater. Sol. Cells* **2011,** *95*, 2194-2199.

6. Savva, A.; Choulis, S. A., Cesium-Doped Zinc Oxide as Electron Selective Contact in Inverted Organic Photovoltaics. *Appl. Phys. Lett.* **2013,** *102*, 233301.

7. Reese, M. O.; Morfa, A. J.; White, M. S.; Kopidakis, N.; Shaheen, S. E.; Rumbles, G.; Ginley, D. S. In *Short-term metal/organic interface stability investigations of organic photovoltaic devices*, Photovoltaic Specialists Conference, 2008. PVSC '08. 33rd IEEE, 11-16 May 2008; 2008; pp 1-3.

8. Paci, B.; Generosi, A.; Albertini, V. R.; Spyropoulos, G. D., Stratakis, E.; Emmanuel Kymakis, Enhancement of Photo/Thermal Stability of Organic Bulk Heterojunction Photovoltaic Devices via Gold Nanoparticles Doping of the Active Layer. *Nanoscale* **2012,** *4,* 7452–7459.

9. Sygletou, M.; Kakavelakis, G.; Paci, B.; Generosi, A.; Kymakis, E.; Stratakis, E., Enhanced Stability of Aluminum Nanoparticle-Doped Organic Solar Cells, *ACS Appl. Mater. Interfaces* **2015,** *7,* 17756−17764.

10. Kakavelakis, G.; Konios, D.; Stratakis, E.; Kymakis, E., Enhancement of the Efficiency and Stability of Organic Photovoltaic Devices via the Addition of a Lithium-Neutralized Graphene Oxide Electron-Transporting Layer *Chem. Mater.* **2014,** *26,* 5988−5993.

11. Yipa, H.-L.; Jen, A. K.-Y., Recent advances in solution-processed interfacial materials for efficient and stable polymer solar cells. *Energy Environ. Sci.,* **2012,** *5,* 5994-6011.

12.    Reese, M. O.; Gevorgyan, S. A.; Jørgensen, M.; Bundgaard, E.; Kurtz, S. R.; Ginley, D. S.; Olson, D. C.; Lloyd, M. T.; Morvillo, P.; Katz, E. A.; Elschner, A.;

Haillant, O.; Currier, T. R.; Shrotriya, V.; Hermenau, M.; Riede, M.; R. Kirov, K.; Trimmel, G.; Rath, T.; Inganäs, O.; Zhang, F.; Andersson, M.; Tvingstedt, K.; Lira-Cantu, M.; Laird, D.; McGuiness, C.; Gowrisanker, S.; Pannone, M.; Xiao, M.; Hauch, J.; Steim, R.; DeLongchamp, D. M.; Rösch, R.; Hoppe, H.; Espinosa, N.; Urbina, A.; Yaman-Uzunoglu, G.; Bonekamp, J.-B.; van Breemen, A. J. J. M.; Girotto, C.; Voroshazi, E.; Krebs, F. C., Consensus Stability Testing Protocols for Organic Photovoltaic Materials and Devices. *Sol. Energy Mater. Sol. Cells* **2011,** *95*, 1253-1267.

13. Norrman, K.; Madsen, M. V.; Gevorgyan, S. A.; Krebs, F. C., Degradation Patterns in Water and Oxygen of an Inverted Polymer Solar Cell. *J. Am. Chem. Soc.* **2010,** *132*, 16883-16892.

14. Giannouli, M.; Drakonakis, V. M.; Savva, A.; Eleftheriou, P.; Florides, G.; Choulis, S. A., Methods for Improving the Lifetime Performance of Organic Photovoltaics with Low-Costing Encapsulation. *ChemPhysChem* **2015**, *16,* 1134-1154.

15. Drakonakis, V. M.; Savva, A.; Kokonou, M.; Choulis, S. A., Investigating Electrodes Degradation in Organic Photovoltaics Through Reverse Engineering Under Accelerated Humidity Lifetime Conditions. *Sol. Energy Mater. Sol. Cells* **2014,** *130*, 544-550.

16. Savva, A.; Georgiou, E.; Papazoglou, G.; Chrusou, A. Z.; Kapnisis, K.; Choulis, S. A., Photovoltaic Analysis of the Effects of PEDOT:PSS-Additives Hole Selective Contacts on the Efficiency and Lifetime Performance of Inverted Organic Solar Cells. *Sol. Energy Mater. Sol. Cells* **2015,** *132*, 507-514.

17. Dupont, S. R.; Voroshazi, E.; Heremans, P.; Dauskardt, R. H., Adhesion Properties of Inverted Polymer Solar Cells: Processing and Film Structure Parameters. *Org. Electron.* **2013,** *14*, 1262-1270.

18. Jorgensen, M.; Norrman, K.; Gevorgyan, S. A.; Tromholt, T.; Andreasen, B.; Krebs, F. C., Stability of Polymer Solar Cells. *Adv. Mater.* **2012,** *24*, 580-612.

19. Stubhan, T.; Li, N.; Luechinger, N. A.; Halim, S. C.; Matt, G. J.; Brabec, C. J., High Fill Factor Polymer Solar Cells Incorporating a Low Temperature Solution Processed $WO_3$ Hole Extraction Layer. *Adv. Energy Mater.* **2012,** *2*, 1433-1438.

20. Zilberberg, K.; Gharbi, H.; Behrendt, A.; Trost, S.; Riedl, T., Low-Temperature, Solution-Processed MoO(x) for Efficient and Stable Organic Solar Cells. *ACS Appl. Mater. Interfaces* **2012,** *4*, 1164-1168.

21. Meyer, J.; Hamwi, S.; Kroger, M.; Kowalsky, W.; Riedl, T.; Kahn, A., Transition Metal Oxides for Organic Electronics: Energetics, Device Physics and Applications. *Adv. Mater.* **2012,** *24*, 5408-5427.

22. Stubhan, T.; Ameri, T.; Salinas, M.; Krantz, J.; Machui, F.; Halik, M.; Brabec, C. J., High Shunt Resistance in Polymer Solar Cells Comprising a $MoO_3$ Hole Extraction Layer Processed From Nanoparticle Suspension. *Appl. Phys. Lett.* **2011,** *98*, 253308.

23. Chen, C. P.; Chen, Y. D.; Chuang, S. C., High-Performance and Highly Durable Inverted Organic Photovoltaics Embedding Solution-Processable Vanadium Oxides as an Interfacial Hole-Transporting Layer. *Adv. Mater.* **2011,** *23*, 3859-3863.

24. Wang, J.-C.; Lu, C.-Y.; Hsu, J.-L.; Lee, M.-K.; Hong, Y.-R.; Perng, T.-P.; Horng, S.-F.; Meng, H.-F., Efficient Inverted Organic Solar Cells Without an Electron Selective Layer. *J. Mater. Chem.* **2011,** *21*, 5723-5728.

25. White, M. S.; Olson, D. C.; Shaheen, S. E.; Kopidakis, N.; Ginley, D. S., Inverted Bulk-Heterojunction Organic Photovoltaic Device Using a Solution-Derived ZnO Underlayer. *Appl. Phys. Lett.* **2006,** *89* (14), 143517.

26. Lloyd, M. T.; Olson, D. C.; Lu, P.; Fang, E.; Moore, D. L.; White, M. S.; Reese, M. O.; Ginley, D. S.; Hsu, J. W. P., Impact of Contact Evolution on the Shelf Life of Organic Solar Cells. *J. Mater. Chem.* **2009,** *19*, 7638-7642.

27. Lloyd, M. T.; Peters, C. H.; Garcia, A.; Kauvar, I. V.; Berry, J. J.; Reese, M. O.; McGehee, M. D.; Ginley, D. S.; Olson, D. C., Influence of the Hole-Transport Layer on the Initial Behavior and Lifetime of Inverted Organic Photovoltaics. *Sol. Energy Mater. Sol. Cells* **2011,** *95*, 1382-1388.

28. Savva, A.; Neophytou, M.; Koutsides, C.; Kalli, K.; Choulis, S. A., Synergistic Effects of Buffer Layer Processing Additives for Enhanced Hole Carrier Selectivity in Inverted Organic Photovoltaics. *Org. Electron.* **2013,** *14*, 3123-3130.

29. Waldauf, C.; Schilinsky, P.; Hauch, J.; Brabec, C. J., Material and Device Concepts for Organic Photovoltaics: Towards Competitive Efficiencies. *Thin Solid Films* **2004,** *451-452*, 503-507.

30. Malliaras, G. G.; Salem, J. R.; Brock, P. J.; Scott, J. C., Photovoltaic Measurement of the Built-in Potential in Organic Light Emitting Diodes and photodiodes. *J. Appl. Phys.***1998,** *84*, 1583.

31. Narioka, S.; Ishii, H.; Yoshimura, D.; Sei, M.; Ouchi, Y.; Seki, K.; Hasegawa, S.; Miyazaki, T.; Harima, Y.; Yamashita, K., The Electronic Structure and Energy Level Alignment of Porphyrin/Metal Interfaces Studied by Ultraviolet Photoelectron Spectroscopy. *Appl. Phys. Lett.* **1995,** *67*, 1899.

32. Soon, Y. W.; Cho, H.; Low, J.; Bronstein, H.; McCulloch, I.; Durrant, J. R., Correlating Triplet Yield, Singlet Oxygen Generation and Photochemical Stability in Polymer/Fullerene Blend Films. *Chem. Commun.* **2013,** *49*, 1291-1293.

33. Voroshazi, E.; Uytterhoeven, G.; Cnops, K.; Conard, T.; Favia, P.; Bender, H.; Muller, R.; Cheyns, D., Root-Cause Failure Analysis of Photocurrent loss in Polythiophene:Fullerene-Based Inverted Solar Cells. *ACS Appl. Mater. Interfaces* **2015,** *7*, 618-23.